\documentclass[preprint,12pt]{elsarticle}

\input{glyphtounicode}
\usepackage{amsthm}
\usepackage{amsmath,amssymb,amsfonts}%
\usepackage{graphicx}
\usepackage{booktabs}
\usepackage{siunitx}
\usepackage[colorlinks,citecolor=red,urlcolor=blue,bookmarks=false,hypertexnames=true]{hyperref}
\usepackage{tabularx}
\usepackage{caption}
\usepackage{subcaption}

\newcommand{\Hc}{H_{\mathrm{c}}}
\newcommand{\Js}{J_{\mathrm{s}}}
\newcommand{\dw}{\delta_{\mathrm{w}}}
\newcommand{\gw}{\gamma_{\mathrm{w}}}
\newcommand{\Lex}{L_{\mathrm{ex}}}
\newcommand{\Dapb}{D_{\mathrm{apb}}}
\newcommand{\Dapbz}{D_{\mathrm{apb},0}}
\newcommand{\gapb}{\gamma_{\mathrm{apb}}}
\newcommand{\Hapb}{H_{\mathrm{apb}}}
\newcommand{\Hgb}{H_{\mathrm{gb}}}
\newcommand{\Hz}{H_{0}}
\newcommand{\degC}{^{\circ}\mathrm{C}}

\journal{Acta Materialia}

\begin{document}
\setlength{\emergencystretch}{3em}

\begin{frontmatter}

\title{Domain-wall energy governs coercivity in ordered and disordered additively manufactured Fe--49Co--2V}

\author[a]{Dennis Boakye\corref{cor}}
\cortext[cor]{Corresponding author}
\ead{dboakye@myumanitoba.ca}
\author[a]{Chuang Deng}

\affiliation[a]{organization={Mechanical Engineering, University of Manitoba},
            addressline={66 Chancellors Cir},
            city={Winnipeg},
            postcode={R3T 2N2},
            state={Manitoba},
            country={Canada}}

\begin{abstract}
	Additively manufactured soft-magnetic Fe--Co is consistently harder than
	wrought material of the same composition, and controlling that excess is the
	central obstacle to printing electrical machines from these alloys. The
	literature attributes it to B2 chemical ordering, untested because the
	anisotropy of the ordered state had not been measured for this alloy and
	because printed material had never been compared with wrought material at
	matched grain size. In this study we supply both. Calibrating on wrought Fe--49Co--2V at a
	known order parameter fixes the ordered-state domain-wall energy and caps
	every pinning channel independent of grain size. Ordering then falls orders of
	magnitude short of the measured excess, and a tenfold change in
	antiphase-domain size leaves coercivity unchanged. What ordering controls is
	the wall energy, and that effect is small and bounded. The excess is retained
	defect content, which moves process optimization from post-build annealing to
	in-build defect control.
\end{abstract}


\end{frontmatter}

%=======================================================================
\section{Introduction}\label{sec:intro}
Additive manufacturing (AM) of soft-magnetic Fe--Co is attractive because the metallurgy that makes these alloys awkward to process conventionally is thermal in origin \cite{nandwana2023disorder,kustas2019controlling}. Near-equiatomic Fe--Co orders into the B2 (CsCl) structure below about $730\,\degC$ \cite{sohn2003magnetic}. The ordered state is brittle and deforms by superdislocation pairs coupled through antiphase boundaries, which suppress cross-slip and promote planar slip and intergranular fracture \cite{marcinkowski1964relationship,sundar2005soft,sourmail2005near}. Wrought production therefore works the alloy in a metastable disordered condition retained by quenching, with up to $3$~wt\% vanadium added to slow the ordering kinetics \cite{sundar2005soft,aykol2010effect}. Cooling rates in laser powder-bed fusion (L-PBF) and directed-energy deposition (DED) exceed $10^{3}\,\mathrm{K\,s^{-1}}$, which prints the alloy under-ordered and allows near-net shapes that cannot be machined from ordered stock \cite{varahabhatla2024influence,nartu2021reducing}.

The magnetics do not follow. AM Fe--Co is consistently harder than wrought material of the same composition, and stays harder after post-build annealing has coarsened the grains \cite{varahabhatla2024influence,nartu2021reducing}. Both of the most complete AM data sets attribute the annealing-induced softening mainly to B2 order. Nartu et al.\ write the coercivity as $\Hc\propto K_1/M_s$ and ascribe the drop to the collapse of $K_1$ on ordering \cite{nartu2021reducing}. Varahabhatla et al.\ attribute their softening to coarse ordered domains alongside grain growth \cite{varahabhatla2024influence}. A companion state-variable theory \cite{boakye2026solidificationcellconfinementdomainwallpinning} sorts the extrinsic coercivity of AM ferromagnets into additive pinning channels,
\begin{equation}
  \Hc = \Hgb + H_{\rho} + H_{\sigma} + H_{\mathrm{ph}} + H_{p},
  \label{eq:masterP1}
\end{equation}
where $\Hgb$ is a grain-boundary (Mager) floor, $H_{\rho}$ a Tr\"auble dislocation term, $H_{\sigma}$ a magnetoelastic microstress term, $H_{\mathrm{ph}}$ a second-phase term and $H_{p}$ a porosity term. That framework holds to within a factor of two across six disordered single-phase printed ferromagnets, but it was noted to leave the ordering alloy Fe--49Co--2V uncaptured \cite{boakye2026solidificationcellconfinementdomainwallpinning}. The obvious repair is a sixth channel carried by antiphase boundaries (APBs), by analogy with the grain-boundary floor.

We build that channel here, and then show its contribution to the hardness of the printed alloys. The reason the question can be settled rather than argued is that the controlling quantity, the magnetocrystalline anisotropy of the ordered state, has been measured. Yu et al.\ \cite{yu1999pinning} report coercivity against grain size for wrought Fe--49Co--2V at an order parameter fixed by neutron diffraction at $S=0.88$. That one data set does three things. Its slope gives the ordered-state wall energy and hence $K_1(S)$. Its intercept places an experimental ceiling on every channel that does not scale with grain size, the APB term included. And it supplies a wrought baseline against which AM material can be compared at matched grain size instead of against a computed floor.

The results are as follows. Evaluated at the measured $K_1(S)$, the APB term comes to between $10^{-8}$ and $10^{-2}$~Oe, and the wrought intercept caps it below $0.79$~Oe independently. What ordering does control is the wall energy $\gw(S)=4\sqrt{AK_1(S)}$, which multiplies every wall-energy pinning term and which we measure falling by a factor between $2.1$ and $4.4$ from $S=0$ to $S=0.88$. The AM residual, isolated against the wrought baseline at matched grain size, runs from $7.6$ to $11.9$ times and is carried by defect content that survives recrystallization.

The rest of the paper is arranged as follows. Section~\ref{sec:channels} sets out the two routes by which ordering can enter a coercivity budget and derives the APB term in both of its regimes. Section~\ref{sec:calib} extracts the calibration from the wrought data. Section~\ref{sec:apbfails} tests the APB hypothesis against it. Section~\ref{sec:residual} isolates the AM residual and asks what carries it. Sections~\ref{sec:pred} give predictions and scope.

%=======================================================================
\section{Two routes by which ordering can enter}\label{sec:channels}

\subsection{Ordering as a wall-energy rescaling}\label{sec:lever}

The grain-boundary floor of Eq.~\eqref{eq:masterP1} follows Mager's
membrane-bowing argument for a $180^{\circ}$ wall pinned between planar
obstacles \cite{mager1952einfluss},
\begin{equation}
  \Hgb = \frac{3\gw}{2\Js\,\bar D_i},
  \qquad \gw = 4\sqrt{A K_1},
  \label{eq:mager}
\end{equation}
with $\bar D_i$ the mean directional grain intercept and $A$ the exchange
stiffness. The prefactor is fixed at $3$ by the bowing geometry. The single
statistical constant of the base framework, $c_1=0.32$, sits on $H_\rho$ and
not on $\Hgb$ \cite{boakye2026solidificationcellconfinementdomainwallpinning}.

Equation~\eqref{eq:mager} is normally applied with $\gw$ treated as an alloy
constant. In an ordering alloy it is not one. The cubic anisotropy of Fe--Co
depends strongly on order: the disordered alloy carries a substantial $K_1$,
whereas ordering drives $K_1$ toward zero, with the crossing composition
moving from about $40$~wt\%~Co when disordered to the stoichiometric
composition when ordered
\cite{hall1959single,sundar2005soft,sourmail2005near}. Since
$\gw\propto\sqrt{K_1}$, every pinning term whose energy scale is the wall
energy gets rescaled by the ordering state. This is the first route, and as we
show below the dominant one. Ordering does not add pinning sites. It changes
what a wall costs.

Two things follow before we look at any data. The rescaling is common-mode: it
multiplies $\Hgb$ and the confinement factor of $H_\rho$ together, so varying
grain size alone cannot isolate it. Separating it requires $S$ to be varied at
fixed microstructure, which is exactly the experiment Yu et al.\ performed
\cite{yu1999pinning}. And it softens the alloy as order develops, which is the
direction both AM data sets report
\cite{varahabhatla2024influence,nartu2021reducing}. The qualitative
observation those papers rest on is therefore reproduced by the rescaling
alone, with no new pinning channel at all.

\subsection{Ordering as a pinning channel}\label{sec:term}

The alternative is that antiphase boundaries pin walls in their own right.
Magnetically an APB is a sheet across which the local easy axis and the
magnitude of the anisotropy are perturbed, and a wall sweeping the ordered
region samples these sheets much as it samples grain boundaries in
Eq.~\eqref{eq:mager}. The pinning energy scale is therefore the
magnetic wall energy $\gw=4\sqrt{AK_1}$, and not the chemical
antiphase-boundary energy $\gapb\simeq0.12$ to $0.16~\mathrm{J\,m^{-2}}$ that
governs superdislocation pair separation and mechanical order-hardening
\cite{marcinkowski1964relationship,aykol2010effect,li2022ordered,sundar2005soft}.
The two differ by roughly two orders of magnitude. We quantify what happens
when they are confused in Section~\ref{sec:energy}.

According to the bowing logic of Eq.~\eqref{eq:mager}, a wall confined between antiphase boundaries of mean intercept $\Dapb$ would depin at
\begin{equation}
  \Hapb^{\mathrm{(bow)}} = \frac{3\gw}{2\Js\,\Dapb}.
  \label{eq:apbbow}
\end{equation}
As with the dislocation channel of Eq.~\eqref{eq:masterP1}, this holds in one
regime only. The controlling quantity is the ratio of the structural scale to
the wall width,
\begin{equation}
  \zeta = \frac{\Dapb}{\dw(S)},
  \qquad \dw(S) = \pi\sqrt{\frac{A}{K_1(S)}},
  \label{eq:zeta}
\end{equation}
and $\dw$ is itself a strong function of $S$ through the same $K_1(S)$ that
sets $\gw$. As order develops the anisotropy collapses and the wall widens.

For $\zeta<1$ the wall overlaps many antiphase domains at once and the pinning
is exchange-averaged, in direct analogy with the diffuse (Tr\"auble) limit of
$H_\rho$ and with random-anisotropy averaging in nanocrystalline magnets
\cite{alben1978random,herzer1990grain,kronmuller2003micromagnetism}. Anisotropy
fluctuations correlated over $\Dapb<\Lex=\sqrt{A/K_1}$ average over
$N=(\Lex/\Dapb)^3$ statistically independent units. Carrying the
self-consistency of the Alben construction through gives an effective
anisotropy $\langle K\rangle=K_1^4\Dapb^{6}/A^{3}$, so that
\begin{equation}
  \Hapb^{\mathrm{(avg)}} = c_5\,\frac{K_1(S)}{\Js}
  \left(\frac{\Dapb}{\Lex}\right)^{\!n}
  = c_5\,\frac{K_1(S)^{4}\,\Dapb^{6}}{A^{3}\,\Js},
  \qquad n=6 .
  \label{eq:apbavg}
\end{equation}
The exponent is not ours to choose. It is the standard three-dimensional
random-anisotropy result \cite{alben1978random,herzer1990grain}, it is the
sixth power the base theory applies to chemical short-range order
[Ref.~\cite{boakye2026solidificationcellconfinementdomainwallpinning}, Eq.~(14)], and it is the form applied independently to
nanocrystalline Fe--Co itself \cite{liu2011effect,jung2003influence}. Neither
is the prefactor: matching the bowing branch at $\Dapb=\dw$ fixes it
analytically at $c_5=6/\pi^{\,n+1}=1.99\times10^{-3}$ for $n=6$
(\ref{app:deriv}). The complete term is switched on $\zeta$ and contains no
constant fitted in this work,
\begin{equation}
  \Hapb =
  \begin{cases}
    \displaystyle \frac{6}{\pi^{\,n+1}}\,\frac{K_1(S)}{\Js}
      \!\left(\frac{\Dapb}{\Lex}\right)^{\!n}, & \Dapb<\dw(S), \\[1.3em]
    \displaystyle \frac{3\gw}{2\Js\,\Dapb}, & \Dapb>\dw(S),
  \end{cases}
  \label{eq:apbfull}
\end{equation}
and the master equation would extend to the ordering alloys as $\Hc=\Hgb+H_{\rho}+H_{\sigma}+H_{\mathrm{ph}}+H_{p}+\Hapb$. Since $\Hapb\to0$ as $S\to0$, that reduces to Eq.~\eqref{eq:masterP1} outside the ordering alloys. Everything in Eq.~\eqref{eq:apbfull} is fixed once $K_1(S)$ and $\Dapb$ are known. 

%=======================================================================
\section{Calibration on wrought Fe--49Co--2V}\label{sec:calib}

\subsection{Experimental data set}\label{sec:yudata}

Yu et al.\ \cite{yu1999pinning} worked with wrought
Co$_{49}$V$_{1.9}$Fe$_{\rm bal}$ sheet, the same commercial composition as the AM material considered below. They varied grain size from $4$ to $21~\mu$m by annealing above the ordering temperature and
cooling at a fixed $90\,\degC$/h. They then measured the long-range order parameter
by neutron diffraction from the integrated intensity ratio $I_{100}/I_{200}$,
obtaining $S=0.88$ for the slow-cooled condition and $S=0$ for brine-quenched
specimens. Saturation magnetization was checked to be unchanged across the
series. This is a controlled variation of grain size at fixed composition and
fixed, independently measured ordering state, which is the combination the AM
data sets cannot provide because grain size and $S$ move together there.

\subsection{Origin of $\gw(S)$ and $K_1(S)$}\label{sec:slope}

\begin{figure}[!ht]
\centering
\includegraphics[width=0.86\linewidth]{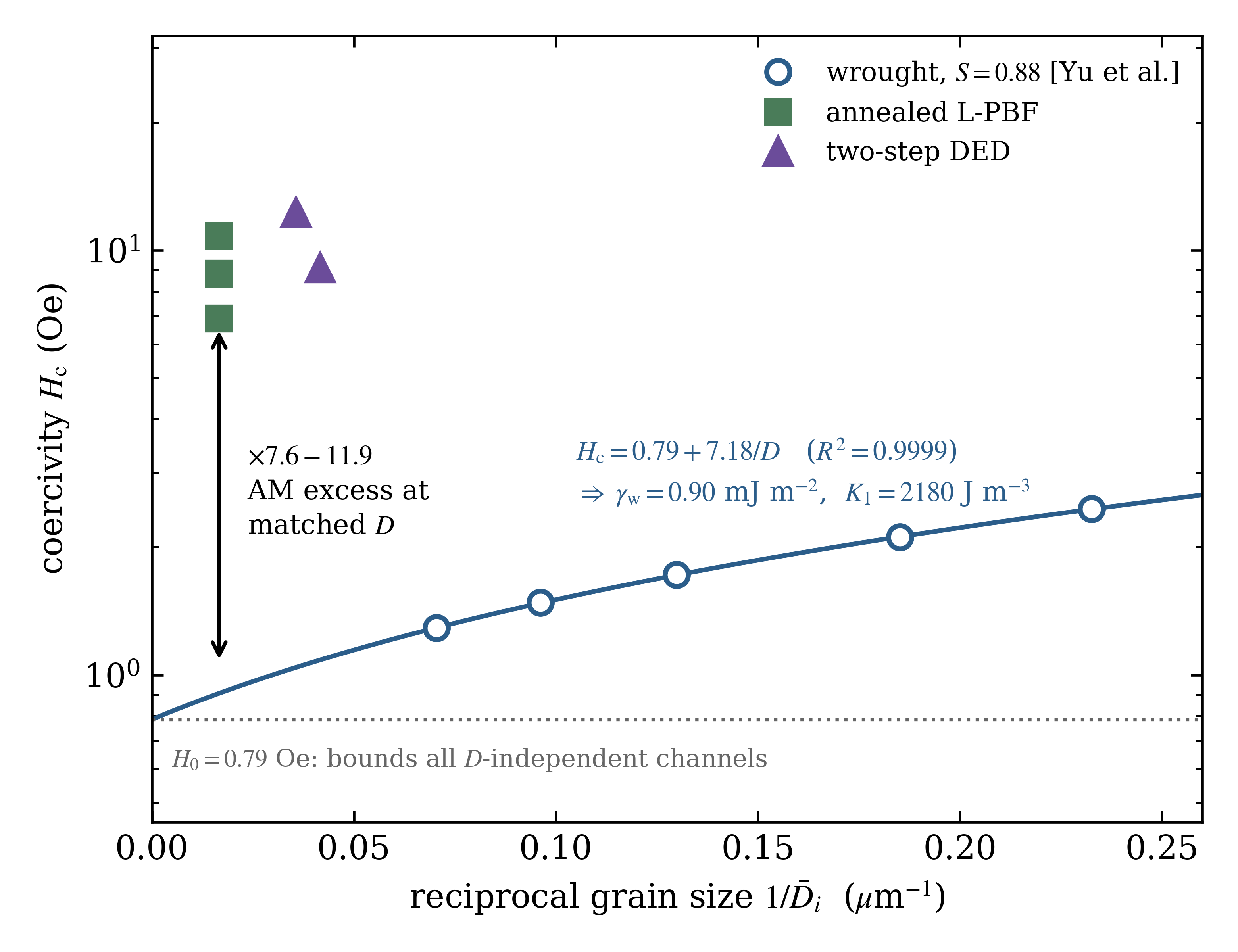}
\caption{Coercivity against reciprocal grain size for wrought Fe--49Co--2V at
an order parameter verified by neutron diffraction at $S=0.88$
\cite{yu1999pinning}. The slope fixes the ordered-state wall energy and hence
$K_1$; the intercept $\Hz$ caps every channel that does not scale with grain
size. Filled symbols are annealed L-PBF \cite{varahabhatla2024influence} and
two-step DED \cite{nartu2021reducing} specimens plotted at their measured
grain sizes. They lie $7.6$ to $11.9$ times above the wrought line at matched
$\bar D_i$.}
\label{fig:calib}
\end{figure}

Their Fig.~9, for the ordered condition annealed above $730\,\degC$, is linear
in $1/\bar D_i$ (Fig.~\ref{fig:calib}). The five points give
\begin{equation}
  \Hc = \Hz + \frac{k}{\bar D_i},
  \qquad \Hz = 0.79~\mathrm{Oe},
  \qquad k = 7.18~\mathrm{Oe}\,\mu\mathrm{m},
  \label{eq:yufit}
\end{equation}
with $R^2=0.9999$. Identifying $k$ with the Mager slope $3\gw/2\Js$ of
Eq.~\eqref{eq:mager} at $\Js=2.35$~T gives
\begin{equation}
  \gw(S{=}0.88) = 0.90~\mathrm{mJ\,m^{-2}},
  \qquad
  K_1(S{=}0.88) = \frac{\gw^2}{16A} = 2.2\times10^{3}~\mathrm{J\,m^{-3}},
  \label{eq:K1meas}
\end{equation}
using $A=2.3\times10^{-11}$~J~m$^{-1}$ \cite{boakye2026solidificationcellconfinementdomainwallpinning}. Three independent routes
agree on this number. Yu et al.'s own fit to the same slopes returns
$1.45\times10^{3}$~J~m$^{-3}$, and torque magnetometry on ternary Fe--Co gives
about $1.5\times10^{3}$~J~m$^{-3}$ \cite{major1988high}, as quoted by
Ref.~\cite{yu1999pinning}. The agreement is a
non-trivial check on the base framework, since the Mager prefactor $3$ and the
identification $\gw=4\sqrt{AK_1}$ together reproduce a measured
coercivity--grain-size slope on this alloy using an anisotropy determined by a
completely different technique.

\subsection{The intercept caps every grain-size-independent channel}
\label{sec:intercept}

The intercept $\Hz=0.79$~Oe is the coercivity this material would retain at
infinite grain size. It is not an artefact of the fit, since it amounts to
$28\%$ of the smallest measured value in the series, and it collects every
contribution that does not scale as $1/\bar D_i$: antiphase-boundary pinning,
residual stress, any $\gamma_2$ precipitate population, and intrinsic terms.
It follows that
\begin{equation}
  \Hapb \le \Hz = 0.79~\mathrm{Oe}
  \label{eq:bound}
\end{equation}
in wrought Fe--49Co--2V at $S=0.88$. This is an experimental ceiling on the
ordering channel that needs no model at all, and it is the single most useful
number in this paper.

\subsection{The wall-energy ratio between ordering states at fixed grain size}\label{sec:leverdata}

Figs.~1 and 6 of Ref.~\cite{yu1999pinning} compare ordered and disordered
specimens through the same annealing series. At $t=2400$~min the two
conditions reach almost the same grain size, about $20$ and $21~\mu$m, with
coercivities of $2.37$ and $1.15$~Oe. At matched grain size the ratio is
therefore a direct measurement of the wall-energy lever,
\begin{equation}
  \frac{\gw(S{=}0)}{\gw(S{=}0.88)} = 2.1 \text{ to } 4.4,
  \qquad
  \gw(S{=}0) = 1.9 \text{ to } 3.9~\mathrm{mJ\,m^{-2}},
  \label{eq:lever}
\end{equation}
where the range spans the assumption that the intercept is common to both
states, which gives the upper bound, or absent, which gives the lower. The
corresponding disordered anisotropy is $K_1(S{=}0)=0.9$ to
$4.1\times10^{4}$~J~m$^{-3}$, a bracket that contains both the
Ref.~\cite{boakye2026solidificationcellconfinementdomainwallpinning} tabulation and Hall's single-crystal values at this
composition \cite{hall1959single}. Ordering therefore softens Fe--49Co--2V by
a factor of two to four through the grain floor alone, at constant
grain size and constant $M_s$. That is the quantitative content of the
qualitative claim both AM studies make.

%=======================================================================
\section{Failure of the antiphase-boundary hypothesis}\label{sec:apbfails}

With $K_1(S)$ measured, Eq.~\eqref{eq:apbfull} can be evaluated rather than
argued about. The ordered-state wall width and exchange length follow at once
from Eq.~\eqref{eq:K1meas},
\begin{equation}
  \dw(S{=}0.88) = 323~\mathrm{nm},
  \qquad \Lex = 103~\mathrm{nm}.
\end{equation}

\subsection{A parameter-free test against two measured domain sizes}\label{sec:modelfree}

Before evaluating anything, the two AM data sets can be played against each
other. Nartu et al.\ report a dark-field antiphase-domain size of about
$100$~nm for the two-step annealed DED alloy \cite{nartu2021reducing}.
Measuring the corresponding dark-field micrograph of the annealed L-PBF
condition [Ref.~\cite{varahabhatla2024influence}, Fig.~5(d)] by line intercept
against its $10$~nm scale bar gives
\begin{equation}
  \Dapb^{\text{L-PBF, annealed}} = 8 \text{ to } 12~\mathrm{nm},
  \label{eq:dapbmeas}
\end{equation}
an order of magnitude finer (\ref{app:arith}). Coarsening kinetics confirm
that independently. Rogers et al.\ measured antiphase-domain growth directly
by dark-field microscopy in wrought Fe--Co--2\%V and found
$\Dapb^2-\Dapbz^2=k(T)t$ with $k=970$~\AA$^2$\,min$^{-1}$ at $823$~K and an
activation energy of $178$~kJ~mol$^{-1}$ \cite{rogers1975electron}.
Extrapolated to $500\,\degC$ that gives $k=108$~nm$^2$\,h$^{-1}$, so a furnace
cool spending of order an hour near the ordering nose, which is what
Ref.~\cite{varahabhatla2024influence}'s $865\,\degC$/4~h treatment amounts to,
produces $\Dapb$ of $8$ to $11$~nm. That is the range we measured. The same
kinetics predict $74$~nm for Nartu et al.'s deliberate $500\,\degC$/50~h hold
against the $100$~nm they report, a $36\%$ discrepancy on a fifty-year
extrapolation. The tenfold separation between the two AM states is not an
artefact of reading a micrograph. It is what their thermal histories require.

The two materials are both well ordered, both have $M_s$ at or near the
commercial Hiperco value, and their grain sizes differ by only a factor of
$2.5$. Their coercivities are $6.9$ to $10.8$~Oe for L-PBF and $9.1$ to
$12.3$~Oe for DED. A tenfold change in antiphase-domain size produces
no systematic change in coercivity at all. The ratio is $0.8$, within
the scatter of either series. Both candidate laws fail this comparison
outright. The bowing floor of Eq.~\eqref{eq:apbbow} predicts the finer-domain
L-PBF material to be ten times harder. The exchange-averaged law of
Eq.~\eqref{eq:apbavg} predicts it to be $10^{6}$ times softer. Neither is what
happens.

This argument uses no value of $K_1$, $A$, $c_5$ or $n$. It needs only two
measured antiphase-domain sizes and two measured coercivities, and on its own
it establishes that whatever sets coercivity in annealed AM Fe--Co is not
indexed by $\Dapb$. The rest of this section quantifies how far short the term
falls, but the conclusion does not rest on it.

\begin{figure}[!ht]
\centering
\includegraphics[width=0.82\linewidth]{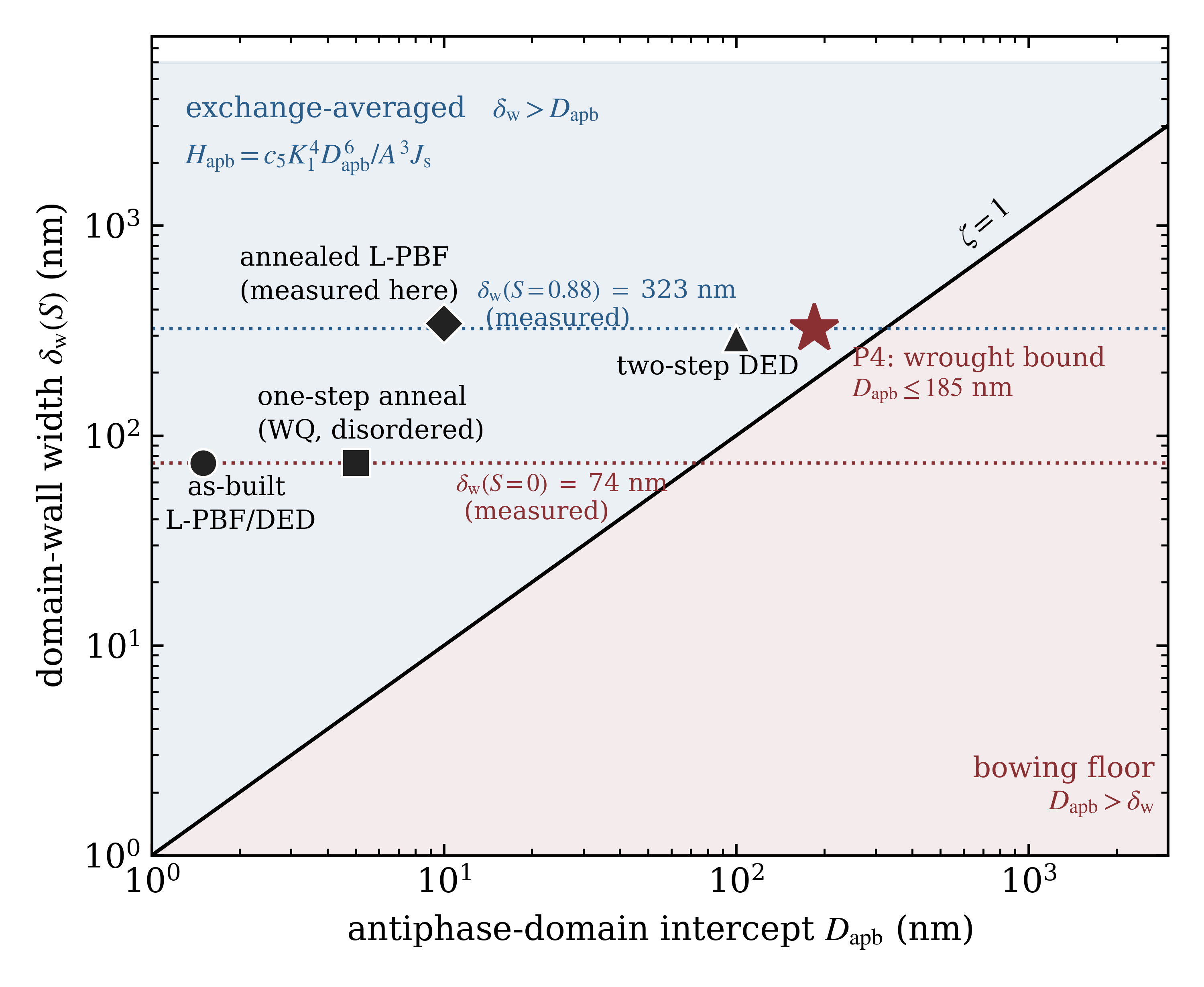}
\caption{Regime map in the $(\Dapb,\dw)$ plane. Both $\dw$ bands are measured
rather than assumed: $\dw(S{=}0.88)=323$~nm from Eq.~\eqref{eq:K1meas} and
$\dw(S{=}0)=74$~nm from the upper branch of Eq.~\eqref{eq:lever}. Every
realized state lies at $\zeta<1$, where $\Hapb\propto K_1^4$ and the term is
negligible. The star marks the ceiling on the wrought antiphase-domain
intercept implied by Eq.~\eqref{eq:bound}, which is prediction P4.}
\label{fig:regime}
\end{figure}

\subsection{The averaging branch falls eight orders short}\label{sec:avgfail}

The measured intercepts put both ordered states well inside the averaging
regime, at $\zeta=0.31$ for DED and $\zeta=0.031$ for L-PBF, so the wall is
three to thirty times wider than the domains it would have to bow between
(Fig.~\ref{fig:regime}). Substituting Eq.~\eqref{eq:K1meas} into
Eq.~\eqref{eq:apbavg} with the continuity-fixed $c_5=6/\pi^{7}$ gives
\begin{equation}
  \Hapb^{\mathrm{(avg)}} =
  \begin{cases}
    2.0\times10^{-2}~\mathrm{Oe}, & \Dapb=100~\mathrm{nm~(DED)},\\[0.3em]
    2.0\times10^{-8}~\mathrm{Oe}, & \Dapb=10~\mathrm{nm~(L\text{-}PBF)}.
  \end{cases}
  \label{eq:apbvalue}
\end{equation}
The annealed AM specimens sit at $6.9$ to $12.3$~Oe. The ordering channel is
therefore between two and eight orders of magnitude too small to be the
residual, and even the larger of the two values is a factor of $40$ below the
wrought intercept that caps it. The reason is the fourth power. Had we instead
inferred $K_1$ by requiring $\Hapb$ to carry the annealed L-PBF
coercivity, we would have got $9.3\times10^{3}$~Jm$^{-3}$, four times the
measured value and six times the torque-magnetometry value. The steepness that
makes Eq.~\eqref{eq:apbavg} a sensitive probe of $K_1$ is exactly what makes
the hypothesis easy to reject once $K_1$ is known independently.

\subsection{The bowing branch is in the wrong regime and too large}
\label{sec:bowfail}

One might instead model the residual as the bowing floor of
Eq.~\eqref{eq:apbbow} evaluated at the measured $\Dapb$, a $1/\Dapb$ law
testable from a single dark-field micrograph. Three objections converge.

\begin{enumerate}
\item \emph{Wrong regime.} Every measured ordered state has $\Dapb<\dw(S)$:
domains of $5$ and $100$~nm \cite{nartu2021reducing}, and $10$~nm here,
against a $323$~nm ordered-state wall. The bowing picture does not apply, by
the same criterion that retires cell confinement at low $K_1$ in
Eq.~\eqref{eq:masterP1}.

\item \emph{Wrong magnitude, and much too steep a trend in $\Dapb$.}
Evaluated at $\Dapb=100$~nm with the measured $\gw$, Eq.~\eqref{eq:apbbow}
returns $71.8$~Oe against $9.1$~Oe measured, eight times too large. At
$\Dapb=5$~nm with the disordered $\gw$ it returns
$2.3\times10^{3}$~Oe against $42.5$~Oe, fifty times too large. The slope fails
with it: across that interval a $1/\Dapb$ law demands a twentyfold softening
whereas $4.7$ times is observed. We stress that the sign of the trend
is not the discriminator, as is sometimes supposed. Bowing softens on
coarsening, and so does the measurement. Both branches of
Eq.~\eqref{eq:apbfull} reproduce that sign, because the one-step to two-step
anneal moves $K_1(S)$ and $\Dapb$ together. Magnitude and slope retire the
bowing floor, not direction.

\item \emph{Wrong energy scale.} Forcing Eq.~\eqref{eq:apbbow} onto the data
with the chemical $\gapb=0.157~\mathrm{J\,m^{-2}}$
\cite{marcinkowski1964relationship} gives $1.3\times10^{4}$~Oe at
$\Dapb=100$~nm, which needs an effective prefactor of about
$7\times10^{-4}$. That factor decomposes exactly: $175$ times from using the
chemical rather than the magnetic energy, multiplied by the eightfold error of
objection~2 from applying the wrong regime.
\end{enumerate}

\subsection{The experimental bound on the ordering channel}\label{sec:boundfail}

Objections~1 to 3 depend on the model. Equation~\eqref{eq:bound} does not.
Wrought Fe--49Co--2V at $S=0.88$, a state at least as well ordered as any AM
specimen and with $M_s$ matching commercial Hiperco, retains at most
$0.79$~Oe once the grain contribution is extrapolated away. Whatever the APB
population contributes in that material, it lies below $0.79$~Oe. Since the AM
excess runs from $6.0$ to $11.3$~Oe (Section~\ref{sec:residual}), ordering can
account for at most $13\%$ of it on the most generous reading, and for $0.3\%$
on the value Eq.~\eqref{eq:apbvalue} actually predicts.

\subsection{The energy-scale distinction}\label{sec:energy}

The most consequential modelling choice above is the use of $\gw$ rather than
$\gapb$. The chemical antiphase energy $\gapb\simeq0.15~\mathrm{J\,m^{-2}}$
\cite{marcinkowski1964relationship,aykol2010effect,li2022ordered,sundar2005soft}
is well measured, but it is the energy of the chemical bond mismatch across
the boundary, and it governs the mechanical order-hardening that motivated
vanadium alloying in the first place. The energy a domain wall exchanges with
an antiphase boundary is the magnetic wall energy, which in the ordered state
is $175$ times smaller. Keeping the two apart is what allows the same
antiphase-domain structure to be strongly hardening mechanically and
negligibly hardening magnetically, which is precisely the situation
Eq.~\eqref{eq:apbvalue} describes.

%=======================================================================
\section{What the AM residual actually is}\label{sec:residual}

\subsection{Isolating it against a matched-grain baseline}\label{sec:excess}

Equation~\eqref{eq:yufit} supplies something the AM literature has lacked: a
wrought reference curve for this composition at known $S$, against which an AM
specimen can be compared at its own grain size. Table~\ref{tab:excess} does
that for every annealed AM specimen with a reported grain size.

\begin{table}[!ht]
\centering
\caption{Annealed AM specimens against the wrought baseline of
Eq.~\eqref{eq:yufit} evaluated at the same grain size. L-PBF data from
Ref.~\cite{varahabhatla2024influence}, Table~1, where all three specimens were
annealed at $865$ to $870\,\degC$ for 4~h and all reached
$\bar D_i=60~\mu$m. DED data from Ref.~\cite{nartu2021reducing}, Tables~1 and
2, with grain sizes from its Figs.~4 and 6.}
\label{tab:excess}
\small
\begin{tabular}{lccccc}
\toprule
condition & $\bar D_i$ & $M_s$ & $\Hc$ & baseline & excess \\
          & ($\mu$m) & (emu\,g$^{-1}$) & (Oe) & (Oe) & \\
\midrule
L-PBF annealed, 6.8~J\,mm$^{-2}$ & 60 & 228 & 6.9  & 0.91 & $7.6\times$ \\
L-PBF annealed, 4.5~J\,mm$^{-2}$ & 60 & 231 & 8.8  & 0.91 & $9.7\times$ \\
L-PBF annealed, 3.4~J\,mm$^{-2}$ & 60 & 228 & 10.8 & 0.91 & $11.9\times$ \\
DED two-step, 189~J\,mm$^{-2}$   & 24 & 242 & 9.1  & 1.09 & $8.4\times$ \\
DED two-step, 47~J\,mm$^{-2}$    & 28 & 257 & 12.3 & 1.04 & $11.8\times$ \\
\bottomrule
\end{tabular}
\end{table}

The excess runs from $7.6$ to $11.9$ times, and it is consistent across two
processes, two feedstocks, two laboratories and a factor of $2.5$ in grain
size. That consistency is itself informative. A residual this reproducible is
a property of AM material as such, not of one particular build.

\subsection{The residual is independent of the ordering state}\label{sec:notordering}

First, the three annealed L-PBF specimens received an identical heat
treatment, reached an identical grain size of $60~\mu$m, and reached
saturation magnetizations of $228$, $231$ and $228$~emu\,g$^{-1}$, which is to
say an identical ordering state, since $M_s$ is the quantity both AM studies
use to track B2 development. Their coercivities are $6.9$, $8.8$ and
$10.8$~Oe. A spread of $1.6$ times at fixed grain size and fixed order cannot
be an ordering effect.

Second, that spread is ordered by laser fluence, and Varahabhatla et al.\
report that as-built kernel-average misorientation rises monotonically as
fluence falls \cite{varahabhatla2024influence}. The excess therefore tracks
the as-built dislocation content through an anneal that recrystallized the
structure and grew the grains twentyfold. What orders the three specimens is
retained defect content, not chemistry.

Third, the DED one-step condition is a null test the ordering hypothesis fails
outright. That specimen was held at $950\,\degC$, above the ordering
temperature, and water-quenched, so it is chemically disordered with
$\Dapb=5$~nm and $\Hapb\sim10^{-6}$~Oe. At $\bar D_i=24~\mu$m its grain floor
is $1.0$~Oe. Its measured coercivity is $42.5$~Oe. A disordered specimen thus
sits $42$ times above its own floor, which is larger than any excess in
Table~\ref{tab:excess}. Whatever produces large residuals in AM Fe--Co
operates with the ordering channel switched off.

\subsection{Retained dislocations, quench stress and $\gamma_2$ precipitation}\label{sec:carriers}

\begin{figure}[!ht]
\centering
\includegraphics[width=0.9\linewidth]{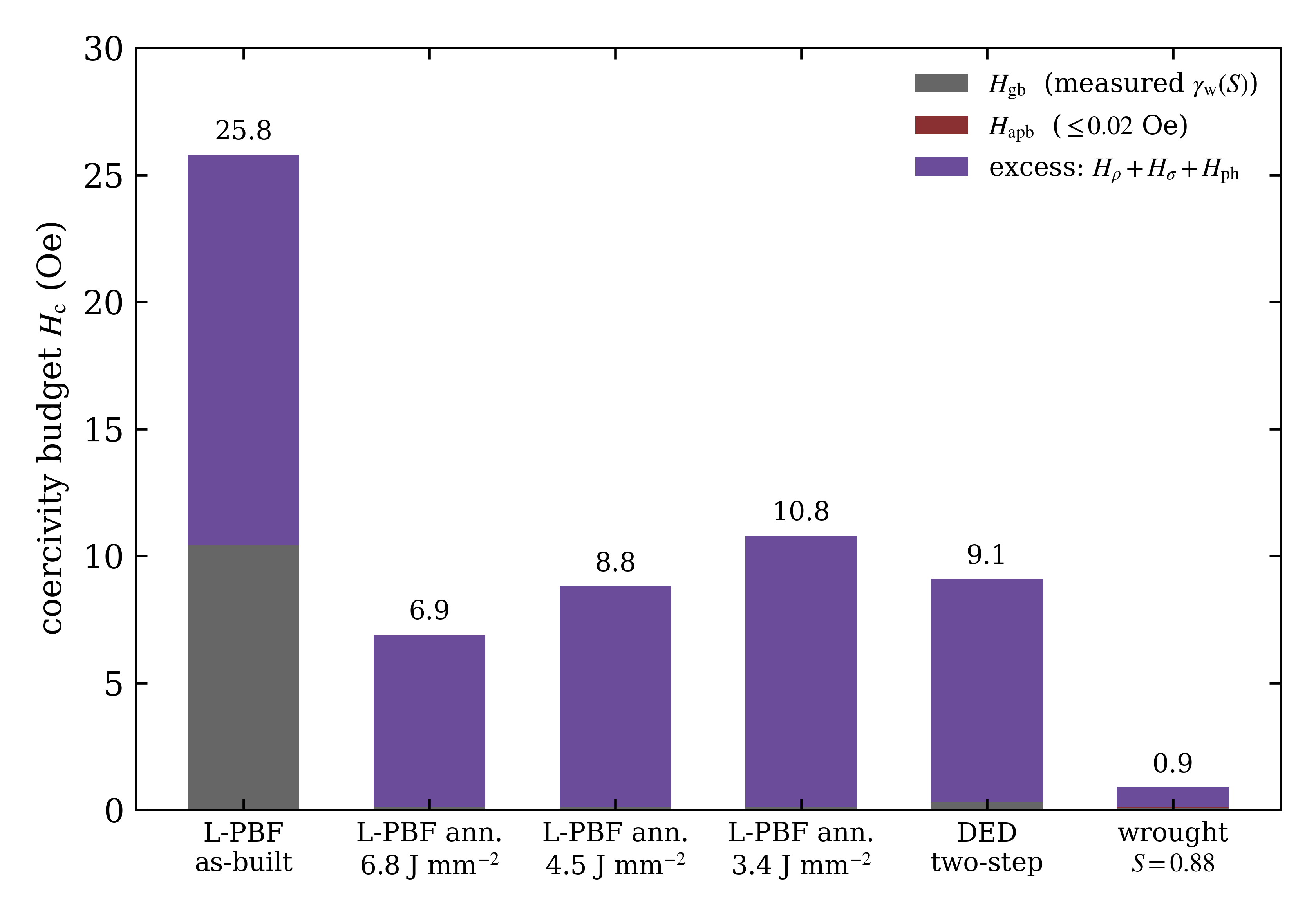}
\caption{Coercivity budget with $\Hgb$ computed from the measured $\gw(S)$ and
$\Hapb$ from Eq.~\eqref{eq:apbvalue}. The ordering channel is invisible at
this scale in every condition. The wrought column shows the intercept $\Hz$,
which caps the same channel experimentally.}
\label{fig:budget}
\end{figure}

Figure~\ref{fig:budget} partitions the budget using the measured $\gw(S)$ and
the capped $\Hapb$. The remainder must be carried by $H_\rho$, $H_\sigma$ and
$H_{\mathrm{ph}}$. Three candidates are consistent with the data and can be
separated experimentally.

\emph{Retained dislocations.} Read through the base-theory dislocation channel
with its frozen $c_1=0.32$, the resolved $\tilde\lambda=\lambda_{111}=2.5
\times10^{-5}$ \cite[Table~4]{sundar2005soft}, together with
$\alpha G_\mu b=6.08$~Pa.m and the confinement factor
$E=\max[1,\sqrt{\lambda_c/2\dw}]$, the annealed excess corresponds to
$\bar\rho\le3.0$ to $7.5\times10^{14}\,\mathrm{m^{-2}}$. The distinction
between bases matters here: the polycrystalline $\lambda_s\simeq6.0$ to
$6.5\times10^{-5}$ often quoted for Fe--Co would lower these densities by a
factor of seven, and in this alloy the two differ sharply, with
$\lambda_{100}=150\times10^{-6}$ against $\lambda_{111}=25\times10^{-6}$
\cite[Table~4]{sundar2005soft}. We take $\lambda_{111}$ because the
dislocation stress field couples through $B_2=-3\lambda_{111}c_{44}$, which is
the same choice made in Ref.~\cite{boakye2026solidificationcellconfinementdomainwallpinning}.
Two features of the result are worth noting. It sits only modestly below the
$\bar\rho\le3.8\times10^{14}\,\mathrm{m^{-2}}$ implied by the as-built excess,
which is hard to reconcile with full recrystallization, except that $4$ to
$11\%$ of the annealed area remains non-recrystallized
\cite{varahabhatla2024influence} and that minority population is exactly where
retained dislocation content would sit. And because $\dw$ widens to $323$~nm
on ordering, $E$ falls to its floor of unity. The cell confinement that
Ref.~\cite{boakye2026solidificationcellconfinementdomainwallpinning} identifies in disordered AM ferromagnets switches
off in the ordered state, leaving the ordered alloy in the diffuse
Tr\"auble limit.

\emph{Quench and residual stress.} The DED one-step residual of $41.5$~Oe
needs only about $100$~MPa through the magnetoelastic channel, given
$\tilde\lambda=2.5\times10^{-5}$. That is an unremarkable level for a water
quench from $950\,\degC$. The subsequent $500\,\degC$/50~h step, which brings
$\Hc$ down to $9.1$~Oe, both develops order and relieves that stress, and the
existing data do not separate the two. Nartu et al.'s attribution of the whole
drop to ordering is therefore not established by their measurements. On the
present accounting the ordering rescaling of $\gw$ can contribute at most a
factor of two to four to the grain term, which is $1$~Oe of a $33$~Oe change.

\emph{$\gamma_2$ precipitation.} Fe--Co--2V precipitates the L1$_2$
$(\mathrm{Co,Fe})_3\mathrm{V}$ phase below about $700\,\degC$ with a C-curve
nose near $550\,\degC$, and vanadium segregates to antiphase boundaries before
precipitation, so $\gamma_2$ forms preferentially on the APB network
\cite{sundar2005soft,ashby1977gamma}. A $600\,\degC$/10~h treatment produces
about $5$~vol\% $\gamma_2$ and raises $\Hc$ by more than $100\%$, with Lorentz
microscopy identifying the precipitates as strong pinning points
\cite{nartu2021reducing}. Both AM anneals cross this field, the L-PBF furnace
cool from $865\,\degC$ and the DED $500\,\degC$/50~h hold, whereas wrought
Hiperco is processed to avoid it. This is an $H_{\mathrm{ph}}$ contribution in
Eq.~\eqref{eq:masterP1}. It is common to all three L-PBF specimens, so it
cannot produce their spread, and it is the most likely carrier of the common
part. It is also an ordering-correlated channel without being an
ordering channel, which is precisely the confound that has led the AM
literature to attribute these residuals to B2 development.

%=======================================================================
\section{Predictions}\label{sec:pred}

\begin{figure}[!ht]
\centering
\includegraphics[width=0.84\linewidth]{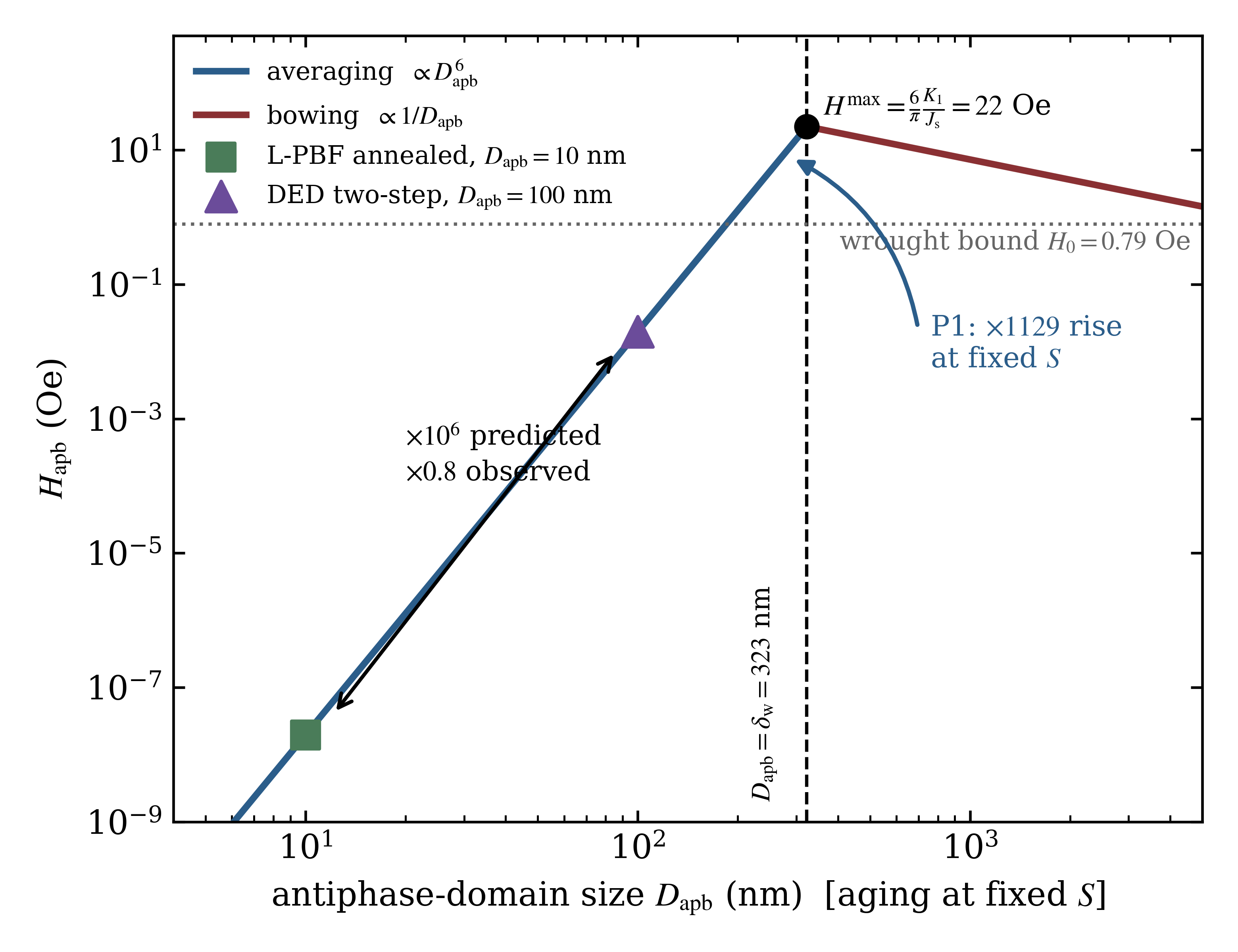}
\caption{Prediction P1, evaluated at the measured $K_1(S{=}0.88)$ rather than
schematically. The two measured states sit far down the averaging branch, and
the dotted line is the experimental ceiling $\Hz$ of Eq.~\eqref{eq:bound},
which the branch respects.}
\label{fig:cross}
\end{figure}

\paragraph{P1: a thousandfold coercivity peak at $\Dapb=\dw(S)$.}
Equation~\eqref{eq:apbfull} is non-monotonic in $\Dapb$ at fixed order. Anneal
a specimen to saturate $S$, which fixes $K_1$, $\dw$ and $\Lex$, then age it
isothermally so that the antiphase structure coarsens at constant $S$. On the
averaging branch $\Hapb$ should rise as $\Dapb^{6}$ while $\Dapb<\dw$, peak at
$\Dapb=\dw(S)$ with height $(6/\pi)K_1/\Js$, and fall as $1/\Dapb$ afterwards
(Fig.~\ref{fig:cross}). Evaluated at the measured $K_1$ the prediction is
extreme, and therefore sharp: coarsening from $100$ to $323$~nm at fixed $S$
should raise $\Hapb$ from $0.02$ to $22$~Oe, a factor of $1.1\times10^{3}$,
taking a specimen from ordering-negligible to ordering-dominated. Nothing else
in the budget moves by three orders of magnitude under an anneal that leaves
the EBSD grain structure alone, so the signature is unambiguous.

The coarsening kinetics of Ref.~\cite{rogers1975electron} fix the cost of the
experiment and with it the temperature at which it should be run. Starting
from $\Dapbz=100$~nm, the deliberately aged DED state, the time to reach
$\Dapb=\dw=323$~nm is

\begin{center}\small
\begin{tabular}{lccccc}
\toprule
$T$ ($\degC$) & 500 & 550 & 600 & 650 & 700 \\
$k$ (nm$^2$\,h$^{-1}$) & 108 & 582 & $2.6\times10^{3}$ & $9.7\times10^{3}$ & $3.2\times10^{4}$ \\
$t$ to $323$~nm & 36 d & 7 d & 37 h & 10 h & 3 h \\
\bottomrule
\end{tabular}
\end{center}

\noindent Both ends of that range are excluded, for opposite reasons. Below
about $600\,\degC$ the specimen sits in the $\gamma_2$ field for weeks. Rogers
et al.\ saw no $\gamma_2$ in ageing at $773$ to $823$~K, which they put down to
short times and to the absence of prior cold work \cite{rogers1975electron},
and both of those also hold for annealed AM material, but their longest hold
was $1000$~min, two orders of magnitude shorter than what is needed here.
Above about $700\,\degC$ the equilibrium order parameter itself starts to fall
as $T_c$ is approached, so $S$ is no longer the saturated value P1 requires.
We note that Rogers et al.'s measurements span $773$ to $823$~K, so the
entries above $550\,\degC$ in the table rest on extrapolating their activation
energy by up to $150$~K; the times quoted there should be treated as
order-of-magnitude guidance for scheduling rather than as predictions. The
window is therefore $650\,\degC$ for about 10\,h:
above the $\gamma_2$ nose, comfortably below $T_c$, at an equilibrium
$S\simeq0.8$ close to the calibrated state, and short enough to be a single
furnace run rather than a campaign. Since $\dw$ scales as $K_1(S)^{-1/2}$, the
crossover should be located from a $K_1$ measurement at the ageing temperature
rather than assumed from the $S=0.88$ value.

What remains difficult is holding $S$ fixed while $\Dapb$ grows, and excluding
$\gamma_2$. The design is to saturate $S$ at the ordering nose, verify by
superlattice intensity, coarsen at $650\,\degC$ with $S$ monitored throughout,
and take paired dark-field $\Dapb$ and vibrating-sample $\Hc$ at each step,
with phase identification to rule out $\gamma_2$. A binary Fe--Co specimen run
alongside removes the $\gamma_2$ field entirely and is the cleaner control, at
the cost of a different $K_1(S)$. If $\Hc$ instead falls monotonically through
the expected crossover with $S$ demonstrably constant,
Eq.~\eqref{eq:apbfull} is falsified and the bowing branch with it.

\paragraph{P2: the annealed residual partitions with as-built cooling rate.}
At fixed grain size and fixed $S$ the residual should scale with retained
defect content rather than with ordering. Section~\ref{sec:notordering} shows
this already holds weakly in the existing flat-grain L-PBF series, where $\Hc$
rises from $6.9$ to $10.8$~Oe as as-built kernel-average misorientation rises.
The prediction proper is that the increment is separable by independent
measurement: $\bar\rho$ from X-ray line broadening or KAM for the $H_\rho$
leg, and $\sigma$ from $\sin^2\psi$ diffraction for the $H_\sigma$ leg, with
$S$ from superlattice intensity held constant across the series so that the
ordering rescaling is excluded. Confirming it means varying cooling rate
deliberately at fixed post-build treatment and decomposing the resulting
change in $\Hc$.

\paragraph{P3: ordering softens through $\gw$, by a bounded factor.} Ageing a
recovered, grain-grown specimen so that $S$ rises at fixed grain size should
lower $\Hc$, but only by a bounded amount. The wall-energy lever of
Eq.~\eqref{eq:lever} rescales $\Hgb$ by a factor between $2.1$ and $4.4$ and
nothing else in the budget by more than that, so the total softening
attributable to ordering cannot exceed the grain-boundary contribution times
that factor. In annealed L-PBF material at $60~\mu$m, $\Hgb\le0.4$~Oe even in
the disordered state, so P3 predicts an ordering-induced softening of at most
$0.3$~Oe against measured annealing drops of $15$ to $33$~Oe. This is the
sharpest available test of the paper's central claim, and it contradicts the
interpretation offered in both AM studies
\cite{varahabhatla2024influence,nartu2021reducing}.

\paragraph{P4: the antiphase-domain intercept of wrought Hiperco.} Inverting
Eq.~\eqref{eq:apbavg} at the ceiling of Eq.~\eqref{eq:bound} gives
\begin{equation}
  \Dapb^{\mathrm{wrought}} \le 185~\mathrm{nm}
  \qquad (\zeta \le 0.57),
  \label{eq:P4}
\end{equation}
since a coarser antiphase structure would push $\Hapb$ above the measured
intercept. One dark-field micrograph of wrought Fe--49Co--2V tests this on
commercially available material with no new processing. A wrought specimen
found at $\Dapb\gg\dw$, in the bowing regime while the AM material sits in the
averaging one, would be the cheapest available falsification of the framework.

%=======================================================================
\section{Discussion}\label{sec:disc}

\subsection{Relation to the base theory}\label{sec:consist}

This work sits inside Eq.~\eqref{eq:masterP1} rather than extending it. We
built the antiphase channel to be the sixth term the base theory appeared to
need, and the outcome is that its contribution is negligible. The residual flagged
in Ref.~\cite{boakye2026solidificationcellconfinementdomainwallpinning} belongs to channels already in the equation. What does
change is that $\gw$ can no longer be treated as an alloy constant in ordering
systems. Both Eq.~\eqref{eq:mager} and the confinement factor of $H_\rho$
carry $\gw(S)=4\sqrt{AK_1(S)}$, and Section~\ref{sec:leverdata} measures that
quantity moving by a factor of two to four within a single alloy. Table~B.1 of
Ref.~\cite{boakye2026solidificationcellconfinementdomainwallpinning} should therefore carry a state label for Fe--49Co--2V:
$\gw=1.9$ to $3.9~\mathrm{mJ\,m^{-2}}$ disordered, and
$0.90~\mathrm{mJ\,m^{-2}}$ at $S=0.88$, the latter measured rather than
assembled from handbook constants.

A second point of contact is the confinement factor. Ref.~\cite{boakye2026solidificationcellconfinementdomainwallpinning}
predicts cellular confinement of dislocation pinning when the wall is narrow
compared with the cell spacing. In ordered Fe--49Co--2V the wall widens to
$323$~nm, $E$ falls to unity, and confinement switches off. The ordered alloy
is thus a natural test of the confinement criterion approached from the
opposite direction to the systems in which it was established.

\subsection{Why the ordering attribution has persisted}\label{sec:why}

Both AM studies attribute their softening to B2 development, and both are
right that ordering and softening are correlated. The correlation has three
sources that the existing data do not separate. There is the genuine but small
$\gw(S)$ rescaling of Section~\ref{sec:lever}. There is the fact that ordering
anneals are also recovery, stress-relief and grain-growth anneals. And there
is the fact that $\gamma_2$ precipitates on the antiphase network, so a
$\gamma_2$ contribution is itself indexed by the ordering state. Telling these
apart needs grain size and $S$ held separately fixed, which is what Yu et al.\
did for wrought material and what P2 and P3 propose for AM material.

\subsection{Scope and limitations}\label{sec:limits}

(i)~The calibration rests on five points digitized from Fig.~9 of
Ref.~\cite{yu1999pinning}. We tabulate the digitized values in
\ref{app:arith} so the fit can be checked or repeated from the original. The
fit quality of $R^2=0.9999$ and the agreement of the extracted $K_1$ with the
authors' own fit to the same slopes, and with the earlier torque determination
they quote \cite{major1988high}, argue that digitization error is not
controlling,
but the intercept in particular would benefit from re-derivation from the
primary data.

(ii)~The exchange stiffness is the one constant not measured on this alloy.
Published values for Fe--Co cluster tightly: $2.0\times10^{-11}$~J~m$^{-1}$
\cite{jung2003influence}, $1.7\times10^{-11}$~J~m$^{-1}$ \cite{liu2011effect},
and the $2.3\times10^{-11}$~J~m$^{-1}$ of Ref.~\cite{boakye2026solidificationcellconfinementdomainwallpinning}. We note that
Ref.~\cite{liu2011effect} attributes its value to
Ref.~\cite{jung2003influence}, which in fact quotes
$2.0\times10^{-6}$~erg~cm$^{-1}$, that is $2.0\times10^{-11}$~J~m$^{-1}$.
Because $A$ enters only through $K_1=\gw^2/16A$ and $\dw=\pi\sqrt{A/K_1}$ once
$\gw$ is fixed by the measured slope, the effect is bounded:

\begin{center}\small
\begin{tabular}{lccccc}
\toprule
$A$ (J\,m$^{-1}$) & $K_1$ (J\,m$^{-3}$) & $K_1/K_1^{\rm torque}$
  & $\dw$ (nm) & $\zeta$ & $\Hapb$ (Oe) \\
\midrule
$1.7\times10^{-11}$ & 2950 & 1.97 & 239 & 0.42 & 0.16 \\
$2.0\times10^{-11}$ & 2510 & 1.67 & 281 & 0.36 & 0.05 \\
$2.3\times10^{-11}$ & 2180 & 1.45 & 323 & 0.31 & 0.02 \\
\bottomrule
\end{tabular}
\end{center}

\noindent Across the full range $\Hapb$ stays below $0.2$~Oe, two orders below
the residual, so the rejection of Section~\ref{sec:apbfails} does not depend
on $A$. What does depend on it is the crossover location, since $\dw$ spans
$239$ to $323$~nm and the P1 peak position carries that spread. We adopt
$A=2.3\times10^{-11}$~J~m$^{-1}$ throughout, both for consistency with
Ref.~\cite{boakye2026solidificationcellconfinementdomainwallpinning} and because it brings the Mager-derived $K_1$ closest to
the torque-magnetometry determination \cite{major1988high}. The residual
factor of $1.45$ between the two is an honest measure of how well the
framework closes on this alloy.

(iii)~Four inputs are carried over or inferred rather than measured on these
specimens, and all of them enter the secondary quantities only. The reference
cell spacing $\lambda_c=0.6~\mu$m and the product
$\alpha G_\mu b=6.08$~Pa\,m are taken from
Ref.~\cite{boakye2026solidificationcellconfinementdomainwallpinning}; no cell
spacing was measured on any specimen considered here, and the elastic
constants behind $\alpha G_\mu b$ are iron-scale generics rather than Fe--Co
values. Both feed the inverted $\bar\rho$ of Section~\ref{sec:carriers},
which should therefore be read as an order of magnitude rather than a
determination. The two-step DED grain sizes are taken equal to the one-step
values on the strength of Ref.~\cite{nartu2021reducing}'s statement that the
second step was designed not to grow grains, which is design intent rather
than a measurement; it shifts the baseline in Table~\ref{tab:excess} by less
than $10\%$ over any plausible range. The wrought grain size used for the
reference column of Fig.~\ref{fig:budget} is the midpoint of the $75$ to
$100~\mu$m quoted in Ref.~\cite{varahabhatla2024influence}, and that column is
illustrative; the quantitative wrought statement in this paper is the
intercept $\Hz$, which is independent of grain size by construction.

(iv)~The disordered wall energy is bracketed rather than measured, because Yu
et al.'s ordered and disordered series are compared at nearly rather than
exactly matched grain size and the intercept of the disordered series is
unknown. Equation~\eqref{eq:lever} is correspondingly wide.

(v)~The antiphase-domain intercept of the annealed L-PBF material,
Eq.~\eqref{eq:dapbmeas}, is measured here from a published micrograph rather
than reported by its authors, and rests on reading the $10$~nm scale bar of
Ref.~\cite{varahabhatla2024influence}, Fig.~5(d). We give the procedure and
its spread in \ref{app:arith}. Because $\Hapb\propto\Dapb^{6}$ the number is
leveraged, but it enters the argument only in the safe direction: the
alternative and larger value of $100$~nm transferred from the DED material
still gives $\Hapb=0.02$~Oe, and the model-free comparison of
Section~\ref{sec:modelfree} uses only the ratio of the two measured
intercepts. The value is corroborated to within a factor of $1.4$ by the
independently measured coarsening kinetics of Ref.~\cite{rogers1975electron}.

(vi)~We do not separate $H_\rho$, $H_\sigma$ and $H_{\mathrm{ph}}$ within the
residual. Section~\ref{sec:carriers} argues that all three are plausibly
present and identifies which is common to the L-PBF series and which produces
its spread, but the partition needs the measurements set out in P2.

(vii)~Two caveats on the source data. Ref.~\cite{varahabhatla2024influence}
quotes $865\,\degC$ in its abstract and $870\,\degC$ in its Table~1, and
reports recrystallized grain sizes as $60~\mu$m in Table~1 and $50$ to
$60~\mu$m in text; we use the tabulated values throughout. The
$M_s=257$~emu\,g$^{-1}$ reported for the $47$~J\,mm$^{-2}$ two-step DED
condition \cite{nartu2021reducing} implies $\Js=2.6$~T at the nominal density,
above the maximum saturation of any Fe--Co composition, and that row should be
treated with caution.

%=======================================================================
\section{Conclusions}\label{sec:concl}

The coercivity penalty of additively manufactured Fe--Co is not a chemical
ordering effect. Both of the most complete data sets on this alloy attribute
the annealing-induced softening to the development of B2 order, and that
reading has gone untested because the ordered-state anisotropy was never
measured and because printed material was never compared with wrought material
at matched grain size. Calibrating on wrought Fe--49Co--2V held at an order
parameter verified by neutron diffraction supplies both. Its
coercivity--grain-size slope fixes the first measured ordered-state wall energy
for this alloy, $\gw=0.90$~mJ~m$^{-2}$, and its intercept caps every
grain-size-independent channel at $0.79$~Oe without recourse to any model.
Evaluated there, an antiphase-boundary pinning term derived in both regimes
and closed by continuity with no fitted constant falls orders of magnitude
short of the residual it was introduced to explain. A tenfold difference in
antiphase-domain size between annealed powder-bed and directed-energy-deposited
material leaves coercivity unchanged to within $20\%$, which rules out any law
indexed on that size using no model parameter at all. Ordering does act on
coercivity, but through the wall energy $\gw(S)=4\sqrt{AK_1(S)}$, which we
measure at fixed grain size to fall by a factor between $2.1$ and $4.4$ and
which can buy at most $0.3$~Oe of softening against annealing drops of $15$ to
$33$~Oe. What remains is retained defect and second-phase content that
survives recrystallization, and it tracks as-built misorientation rather than
the degree of order. The practical consequence is that post-build heat
treatment cannot recover wrought coercivity in this alloy, and that
optimization belongs in the build. Four predictions set out above test each of
these claims independently.

%=======================================================================
\section*{CRediT authorship contribution statement}
\textbf{Dennis Boakye}: Writing -- review and editing, Writing -- original
draft, Visualization, Validation, Methodology, Investigation, Formal analysis,
Data curation. \textbf{Chuang Deng}: Writing -- review and editing,
Supervision, Resources, Project administration, Funding acquisition,
Conceptualization.

\section*{Declaration of competing interest}
The authors declare no competing financial interests.

\section*{Data availability}
All data used are from the cited public sources. The digitized calibration
data and the full arithmetic of Sections~\ref{sec:calib} to
\ref{sec:residual} are reproduced in \ref{app:arith}.

\section*{Acknowledgements}
This research was supported by the NSERC Alliance International Catalyst
(ALLRP 592696-24), Canada.

%=======================================================================
\appendix
\section{Bowing branch and the crossover prefactor}\label{app:deriv}

The bowing branch follows the same overlap-integral argument as the grain
floor. A $180^{\circ}$ wall of width $\dw$ and areal energy $\gw$ traversing a
planar antiphase boundary picks up an excess energy that rises from zero to a
plateau over a traversal distance of order $\dw$. Taking the maximum restoring
gradient, dividing by $2\Js$ and evaluating for a mean boundary spacing
$\Dapb$ gives Eq.~\eqref{eq:apbbow}, with the geometric prefactor $3$
inherited from the grain-floor derivation of Ref.~\cite{boakye2026solidificationcellconfinementdomainwallpinning}.

At $\Dapb=\dw(S)$ the two branches must match. Using $\gw=4\sqrt{AK_1}$,
$\dw=\pi\sqrt{A/K_1}$ and $\Lex=\sqrt{A/K_1}$, so that $\dw/\Lex=\pi$ exactly,
\begin{align}
  \Hapb^{\mathrm{(bow)}}\big|_{\Dapb=\dw}
    &= \frac{3\cdot 4\sqrt{AK_1}}{2\Js\,\pi\sqrt{A/K_1}}
     = \frac{6}{\pi}\,\frac{K_1}{\Js}, \\
  \Hapb^{\mathrm{(avg)}}\big|_{\Dapb=\dw}
    &= c_5\,\frac{K_1}{\Js}\left(\frac{\dw}{\Lex}\right)^{n}
     = c_5\,\frac{K_1}{\Js}\,\pi^{n},
\end{align}
so continuity fixes
\begin{equation}
  c_5 = \frac{6}{\pi^{\,n+1}},
  \label{eq:c5}
\end{equation}
which for $n=6$ gives $c_5=1.99\times10^{-3}$. The peak coercivity is then
$\Hapb^{\max}=(6/\pi)K_1(S)/\Js$, independent of $\Dapb$, $A$ and $c_5$ alike.
Written out, $\Hapb^{\mathrm{(avg)}}=c_5K_1^{4}\Dapb^{6}/A^{3}\Js$ is the
Alben--Herzer effective anisotropy $\langle K\rangle=K_1^4d^6/A^3$
\cite{alben1978random,herzer1990grain} divided by $\Js$ and scaled by $c_5$.
It is therefore the same construction the base theory applies to chemical
short-range order [Ref.~\cite{boakye2026solidificationcellconfinementdomainwallpinning}, Eq.~(14)], with the short-range-order
domain size replaced by the antiphase intercept. Equation~\eqref{eq:c5} is
what removes the last adjustable quantity from the ordering term.

\section{Calibration arithmetic}\label{app:arith}

\paragraph{Digitized calibration data.} Read from Fig.~9 of
Ref.~\cite{yu1999pinning}, for wrought Fe--49Co--2V annealed above
$730\,\degC$, cooled at $90\,\degC$/h, at $S=0.88$ by neutron diffraction:

\begin{center}\small
\begin{tabular}{lccccc}
\toprule
$\bar D_i$ ($\mu$m) & 4.3 & 5.4 & 7.7 & 10.4 & 14.2 \\
$\Hc$ (Oe)          & 2.46 & 2.11 & 1.72 & 1.48 & 1.29 \\
\bottomrule
\end{tabular}
\end{center}

Least squares against $1/\bar D_i$ gives $\Hz=0.786$~Oe,
$k=7.18~\mathrm{Oe}\,\mu\mathrm{m}$ and $R^2=0.9999$.

\paragraph{From slope to anisotropy.} With $1$~Oe~$=79.577$~A~m$^{-1}$ the
slope is $k=5.716\times10^{-4}$~A. Setting $k=3\gw/2\Js$ at $\Js=2.35$~T gives
$\gw=8.96\times10^{-4}$~J~m$^{-2}$, and $K_1=\gw^2/16A$ with
$A=2.3\times10^{-11}$~J~m$^{-1}$ gives $K_1=2.18\times10^{3}$~J~m$^{-3}$,
whence $\dw=\pi\sqrt{A/K_1}=323$~nm and $\Lex=\sqrt{A/K_1}=103$~nm.

\paragraph{The ordering lever.} At $t=2400$~min in Figs.~1 and 6 of
Ref.~\cite{yu1999pinning} the disordered and ordered specimens reach
$\bar D_i\simeq20$ and $21~\mu$m with $\Hc=2.37$ and $1.15$~Oe. The ratio
$2.37/1.15=2.06$ assumes zero intercept in both states, while
$(2.37-\Hz)/(1.15-\Hz)=4.35$ assumes the intercept is common. These bracket
$\gw(S{=}0)=1.85$ to $3.90~\mathrm{mJ\,m^{-2}}$ and $K_1(S{=}0)=0.93$ to
$4.13\times10^{4}$~J~m$^{-3}$.

\paragraph{The ordering term.} With $c_5=6/\pi^7=1.99\times10^{-3}$,
$K_1=2.18\times10^{3}$~J~m$^{-3}$, $\Dapb=100$~nm,
$A=2.3\times10^{-11}$~J~m$^{-1}$ and $\Js=2.35$~T,
$\Hapb=c_5K_1^4\Dapb^6/(A^3\Js)=1.57$~A~m$^{-1}=0.0197$~Oe. The bowing branch
at the same $\Dapb$ is $3\gw/(2\Js\Dapb)=5.72\times10^{3}$~A~m$^{-1}=71.8$~Oe.
Inverting the averaging branch at $\Hapb=\Hz$ gives
$\Dapb=100\,(\Hz/0.0197)^{1/6}$~nm~$=185$~nm, which is Eq.~\eqref{eq:P4}.

\paragraph{Antiphase-domain intercept of the annealed L-PBF alloy.} Measured
from Fig.~5(d) of Ref.~\cite{varahabhatla2024influence}. The labelled scale
box spans $168$~px for $10$~nm, so the micrograph field is $28\times23$~nm. We
Gaussian-smoothed the image to suppress dark-field speckle, binarized it at
its median, and counted the mean linear intercept along both image axes,
obtaining $7.8$, $8.4$ and $11.3$~nm at smoothing lengths of $0.8$, $1.0$ and
$1.5$~nm. The radial autocorrelation of the smoothed image falls to zero at
$8.5$~nm. We adopt $\Dapb=10$~nm with a spread of $8$ to $12$~nm. The as-built
micrograph, Fig.~5(b), carries an identical $168$~px scale box, which confirms
the calibration. Its contrast sits at finer scales than the annealed
condition: on equal $15.5$~nm regions of interest and identical processing,
the annealed image carries $37\%$ of its spectral power in the $5$ to $15$~nm
band against $16\%$ for the as-built, while the as-built carries $28\%$ below
$5$~nm against $14\%$ annealed. We therefore assign
$\Dapb^{\rm as\text{-}built}=2$ to $5$~nm, a coarsening ratio of $2$ to $5$
across the anneal. Both values put $\Hapb$ below $10^{-9}$~Oe, so neither is
load-bearing; the ratio matters only for the P1 ageing estimate.

\paragraph{Dislocation densities.} The base-theory channel
$H_\rho=c_1(3\tilde\lambda/2\Js)\alpha G_\mu b\sqrt{\bar\rho}\,E$ with
$c_1=0.32$, $\tilde\lambda=\lambda_{111}=2.5\times10^{-5}$,
$\alpha G_\mu b=6.08$~Pa\,m and
$E=\max[1,\sqrt{\lambda_c/2\dw}]$ at $\lambda_c=0.6~\mu$m gives $E=2.01$ for
$\dw(S{=}0)=74$~nm and $E=1$, the diffuse limit, for $\dw(S{=}0.88)=323$~nm.
Inverting for the excesses of Fig.~\ref{fig:budget} gives
$\bar\rho\le3.8\times10^{14}$~m$^{-2}$ as-built and $3.0$ to
$7.5\times10^{14}$~m$^{-2}$ annealed.

%=======================================================================
\scriptsize
\bibliographystyle{elsarticle-num}
\biboptions{sort&compress}
\bibliography{references}

\end{document}